\documentclass[twocolumn,amsmath,amssymb,prl,superscriptaddress,longbibliography]{revtex4-1}

\setcitestyle{super}
\usepackage{bm}

\usepackage[final,letterspace=-60]{microtype}
\usepackage[colorlinks=true, linkcolor=blue, urlcolor=blue,  citecolor=blue, anchorcolor=blue]{hyperref}

\usepackage{mathpazo} 

\usepackage{graphicx}
\usepackage{dcolumn}
\usepackage{bm}
\usepackage{helvet}

\usepackage{caption}
\DeclareCaptionLabelFormat{plain}{Figure #2 $\boldsymbol |$ }

\begin{document}

\title{Self-healing of space-time light sheets}

\author{H. Esat Kondakci}
\email{esat@knights.ucf.edu}
\author{Ayman F. Abouraddy}
\email{raddy@creol.ucf.edu}
\affiliation{CREOL, The College of Optics \& Photonics, University of Central Florida, Orlando, Florida 32816, USA}

\begin{abstract} \noindent 	
Space-time wave packets are diffraction-free, dispersion-free pulsed beams whose propagation-invariance stems from correlations introduced into their spatio-temporal spectrum. We demonstrate here experimentally and computationally that space-time light sheets exhibit self-healing properties upon traversing obstacles in the form of opaque obstructions. The unscattered fraction of the wave packet retains the spatio-temporal correlations and thus propagation-invariance is maintained. The scattered component does not satisfy the requisite correlation and thus undergoes diffractive spreading. These results indicate the robustness of ST wave packets and their potential utility for deep illumination and imaging in scattering media such as biological tissues.
\end{abstract}

\maketitle


\noindent 

Diffraction-free beams are fields whose transverse profiles are independent of the axial coordinate upon free propagation \cite{Durnin87PRL}. A wide range of monochromatic fields display this behavior, including Bessel, Mathieu, Weber, and self-accelerating Airy beams \cite{Levy16PO}. In addition to their numerous salutary features, diffraction-free beams have `self-healing' characteristics; that is, despite distortion in the field distribution after traversing an obstruction, the beam nevertheless approximately retrieves its original distribution after freely propagating a finite distance \cite{Broky08OE,Hu13PRA,Bongiovanni16OE}.

Propagation-invariant \textit{pulsed} beams or wave packets -- i.e., those that are diffraction-free \textit{and} dispersion-free -- have had an extensive history starting from theoretically proposed solutions such as Brittingham's focus-wave modes \cite{Brittingham83JAP}, Mackinnon's wave packet \cite{Mackinnon78FP}, X-waves \cite{Lu92IEEEa}, and Bessel-pulse X-waves \cite{Sheppard19JOSAA,Christodoulides04OE}, and subsequent experimental realizations \cite{Saari97PRL} (see \cite{Turunen10PO,FigueroaBook14} for reviews). Such wave packets propagate self-similarly with no change in the field spatial or temporal profiles. The propagation invariance of these pulsed fields is a consequence of the spatio-temporal spectrum incorporating tight correlations between the \textit{spatial} frequencies that determine the field spatial profile and the \textit{temporal} frequencies or wavelengths that determine the pulse linewidth. We thus term all such pulsed beams `space-time' (ST) wave packets \cite{Kondakci16OE}. We have recently demonstrated the synthesis and characterization of a wide range of $(1+1)$D ST optical sheets in which the field along one transverse dimension is uniform \cite{Kondakci17NP}, including non-accelerating ST Airy wave packets that instead accelerate in the local space-time frame of the propagating pulse \cite{Kondakci18PRL}. Additionally, several theoretical treatments of this class of wave packets have appeared very recently \cite{Parker16OE,Wong17ACSP2,Porras17OL,Efremidis17OL,SaintMarie17Optica,Porras18}.

The question we pose in this Letter is the following: are such propagation-invariant ST wave packets endowed with self-healing characteristics similar to those displayed by monochromatic diffraction-free beams? One is initially inclined to suppose the answer in the affirmative from the analogous self-similar propagation of both classes of fields. However, the origins of their propagation invariance are distinct. For monochromatic diffraction-free beams, the transverse spatial profiles are eigensolutions of the Helmholtz equation in various coordinate systems. For ST wave packets, propagation invariance is a result of spectral correlations between the spatial and temporal degrees of freedom of the field. 

Here we demonstrate experimentally and computationally that ST wave packets are indeed endowed with self-healing characteristics. We introduce 1D opaque spatial obstructions (widths 6, 10, and 20 $\mu$m) in the path of diffraction-free ST light sheets (of widths in the range $7-23$~$\mu$m) and observe their subsequent axial evolution. We find that the sheets self-heal within a few millimeters, as predicted by a computational model. Our observations indicate the robustness of this new class of pulsed beams and their potential utility for imaging in turbid media such as biological tissues.

\begin{figure}[t!]
	\centering
	\includegraphics[width=8cm]{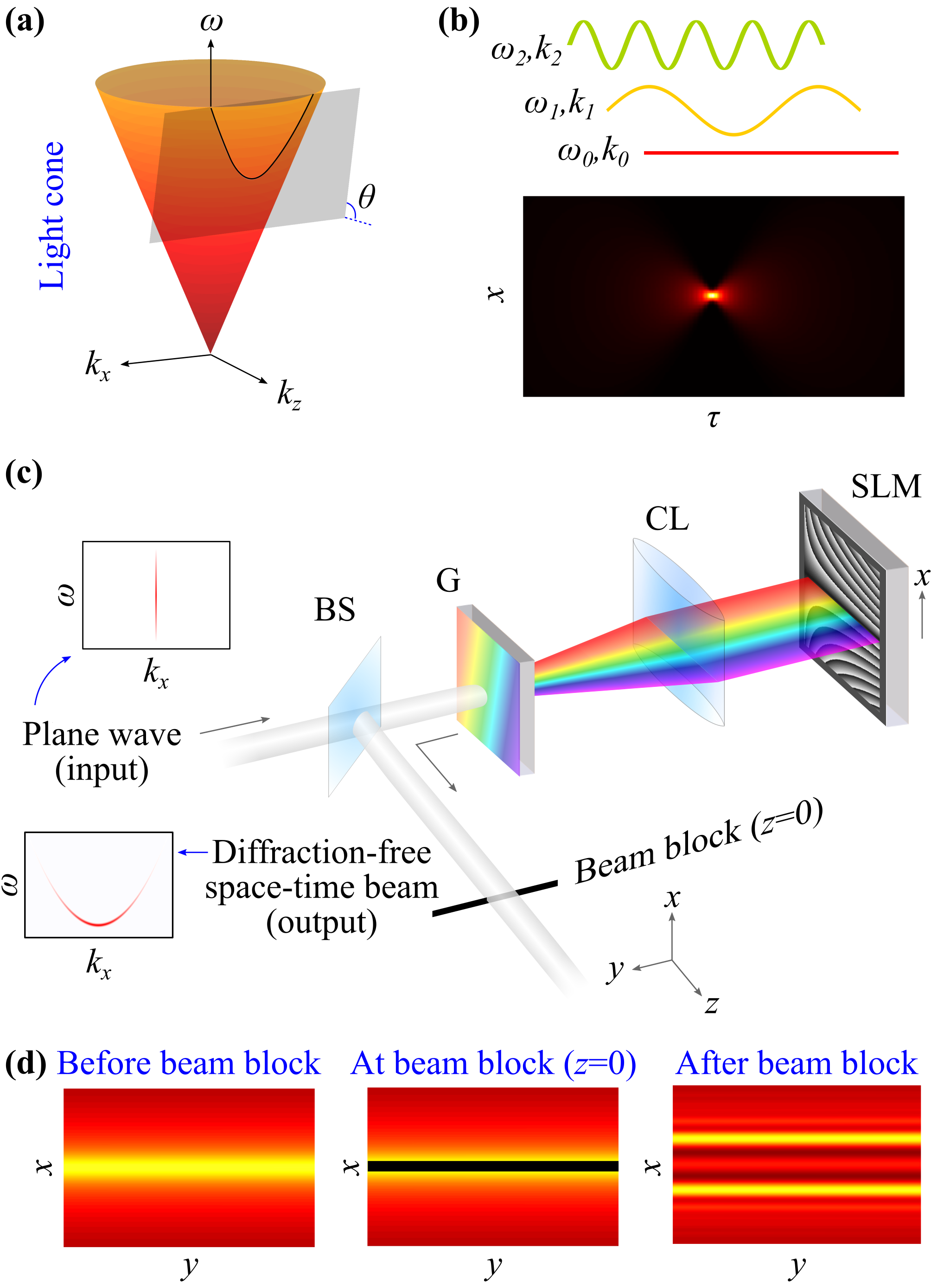}
	\caption{{(a) Concept of ST wave packets. The spatio-temporal spectral locus of such wave packets lie along the intersection of the light-cone $k_{x}^{2}+k_{z}^{2}=(\omega/c)^{2}$ with a spectral hyperplane tilted an angle $\theta$ with respect to the horizontal $(k_{x},k_{z})$-plane. (b) Intensity $|E(x,z,\tau)|^{2}$ of a generic ST wave packet (calculated with a Gaussian spatio-temporal spectrum and $\theta=90^{\circ}$) at a fixed axial position $z$; $\tau$ is time in the moving pulse frame. Higher \textit{temporal} frequencies $\omega$ are assigned to larger \textit{spatial} frequencies. (c) Schematic of the setup for synthesizing ST wave packets. BS: Beam splitter; G: diffraction grating; CL: cylindrical lens; SLM: spatial light modulator. (d) Time-averaged intensity of the transverse spatial profile $I(x,y,z)=\int dt|E(x,y,z;t)|^{2}$ at three \textit{axial} positions: $z<0$ before the obstacle; $z=0$ immediately after the opaque obstruction; and $z>0$ after propagation beyond the obstacle where the scattered component and the emerging self-healed components are observed.}}
	\label{Fig:Concept}
\end{figure}

The ST wave packets in the form of light sheets that we study here are synthesized in such a way that each spatial-frequency pair $\pm k_{x}$ (the transverse component of the wave vector along the $x$-axis) is associated with one temporal (angular) frequency $\omega$, which also implies in turn a correlation between the axial wave number $k_{z}$ (along the propagation axis $z$) and $\omega$. We assume that the field is uniform along the second transverse dimension $y$; i.e., $k_{y}=0$. By enforcing a linear relationship of the form $\omega/c=k_{\mathrm{o}}+v_{\mathrm{g}}(k_{z}-k_{\mathrm{o}})$ between $k_{z}$ and $\omega$, the envelope of the wave packet $E(x,z,t)=\psi(x,z,t)e^{i(k_{\mathrm{o}}z-\omega_{\mathrm{o}}t)}$ undergoes the transformation $\psi(x,z,t)\rightarrow\psi(x,z-v_{\mathrm{g}}t)$, which enforces rigid translation of the wave packet along the propagation axis. The reduced functional-dimensionality of the field is associated with a reduced dimensionality spatio-temporal spectrum. Instead of a 2D patch on the surface of the light-cone for a typical pulsed beam (one dimension for $\omega$ and the other for $k_{x}$), ST wave packets are associated with reduced-dimensionality trajectories \cite{Kondakci17NP}. Specifically, the locus of the spatio-temporal spectrum is at the intersection of the light-cone with a spectral hyperplane parallel to the $k_{x}$-axis as shown in Fig.~\ref{Fig:Concept}(a). By virtue of correlating each spatial frequency $|k_{x}|$ to a temporal frequency $\omega$, a propagation-invariant wave packet is formed [Fig.~\ref{Fig:Concept}(b)].

The experimental arrangement for the synthesis of ST light sheets relies on introducing the spectral spatio-temporal correlations via a spatial light modulator (SLM; Hamamatsu X10468-02) as shown schematically in Fig.~\ref{Fig:Concept}(c). Starting with a pulsed plane wave from a mode-locked femtosecond Ti:Sa laser (Spectra Physics, Tsunami), a reflective diffraction grating (1200 lines/mm, area $25\times25$~mm$^2$) spreads the spectrum, followed by a collimating cylindrical lens. The spectrum then impinges on the reflective SLM that displays a 2D phase distribution $\Phi(x,y)$, where $y$ refers to the direction along which the spectrum is spread. In contrast to traditional pulse-shaping where a 1D phase distribution modulates the spatially dispersed pulse spectrum, here the axis \textit{orthogonal} to spread spectrum (i.e., the $x$ direction) implements a linear phase at each position $y$ (and hence each wavelength) whose slope corresponds to the spatial frequency $\pm k_{x}$ we assign to that particular wavelength. The reflected wave front passes through the cylindrical lens and then the diffraction grating, which combines the wavelengths and reconstitutes the pulse. This simultaneously leads to the overlap of the spatial frequencies and thus the formation of the beam in the same step [see Fig.~\ref{Fig:Concept}(c)]. We make use of an SLM here to introduce the spatio-temporal correlations, but refractive phase plates can be alternatively exploited \cite{Kondakci18OE}. The ST wave packet has a temporal bandwidth of $\Delta\omega$ (or $\Delta\lambda$ in terms of wavelengths) and a spatial bandwidth of $\Delta k_{x}$, which are not independent but are related through the tilt angle $\theta$ of the spectral hyperplane [Fig.~\ref{Fig:Concept}(a)]. In the ideal case, a delta-function correlation between $k_{x}$ and $\omega$ guarantees propagation-invariance indefinitely, but this requires an infinitely wide grating and illumination area, thus entailing an infinite energy per pulse. In practice, finite apertures induce a small spectral uncertainty $\delta\lambda$ in the correlation between $\lambda$ and $k_{x}$, which limits the propagation-invariant distance \cite{Kondakci16OE}.

The result is a sheet of light whose transverse profile is uniform along $y$ and narrow along $x$ [Fig.~\ref{Fig:Concept}(d), left panel]. We refer henceforth to the full-width at half-maximum (FWHM) width along $x$ as $\Delta x$, which is determined by the spatial bandwidth $\Delta k_{x}$ encoded on the SLM. To test the self-healing properties of the ST light sheets synthesized here, we place an obstructive object in the form of an opaque line aligned along the center of the light sheet at the plane $z=0$ [Fig.~\ref{Fig:Concept}(d), central panel]. The obstacles we use in our experiments are in the form of extended, straight opaque lines of chrome of varying width on soda lime glass substrates. The obstacles were all produced by lithographically patterning a 5-inch square blank glass substrate with a 100-nm-thick layer of chrome and a 50~nm-thick layer of photoresist, followed by dicing the substrate into 1-inch square samples using a glass cutter. After traversing this obstacle, the field evolves [Fig.~\ref{Fig:Concept}(d), right panel], and we expect that the original form of the field [Fig.~\ref{Fig:Concept}(d), left panel] is regained after free axial propagation.

We first examine theoretically the self-healing characteristics of an ST wave packet with an opaque uniform obstacle of width $d$ placed in its path. The field after traversing such an obstacle can be expressed at the sum of two terms,
 \begin{eqnarray}\label{Eq:MainEquation}
 &E&(x,z,t)=e^{ik_{\mathrm{o}}z}\int dk_{x}F(k_{x})e^{i(k_{x}x-g(k_{x})t)}\nonumber\\
 &+&W\iint dk_{x}d\omega F(g^{-1}(\omega))\mathrm{sinc}\left(\frac{W(k_{x}-g^{-1}(\omega))}{2\pi}\right)e^{i(k_{x}x+k_{z}z-\omega t)},\nonumber\\\vspace{-2mm}
 \end{eqnarray}
where the first term is the unscattered component of the ST propagation-invariant field that maintains the correlation $\omega=g(k_{x})$ between $k_{x}$ and $\omega$, the second term is the scattered field in which this correlation is no longer retained, $\tilde{E}(k_{x})$ is the spatial spectrum of the ST wave packet, and $\mathrm{sinc}(x)=\tfrac{\sin{\pi x}}{\pi x}$. The form of $g(k_{x})$ depends on the tilt angle $\theta$ in Fig.~\ref{Fig:Concept}(a). 

The self-healing properties of ST wave packets can be understood as follows. In a ST wave packet, each spatial frequency $k_{x}$ is correlated to a temporal frequency $\omega$. Upon incidence on an object, the scattered portion (second term in Eq.~\ref{Eq:MainEquation}) consists of components for which this correlation is violated. Although $\omega$ is invariant, it is now associated with a range of spatial frequencies $k_{x}\rightarrow\{k_{x}'\}$ that do not satisfy the original ST correlation. The scattered component is thus no longer propagation-invariant and instead undergoes the usual diffractive spreading. The unscattered component (first term in Eq.~\ref{Eq:MainEquation}) emerges intact when the scattered component has diffracted after a few Rayleigh lengths.

\begin{figure}[b!]
	\centering
	\includegraphics[width=8cm]{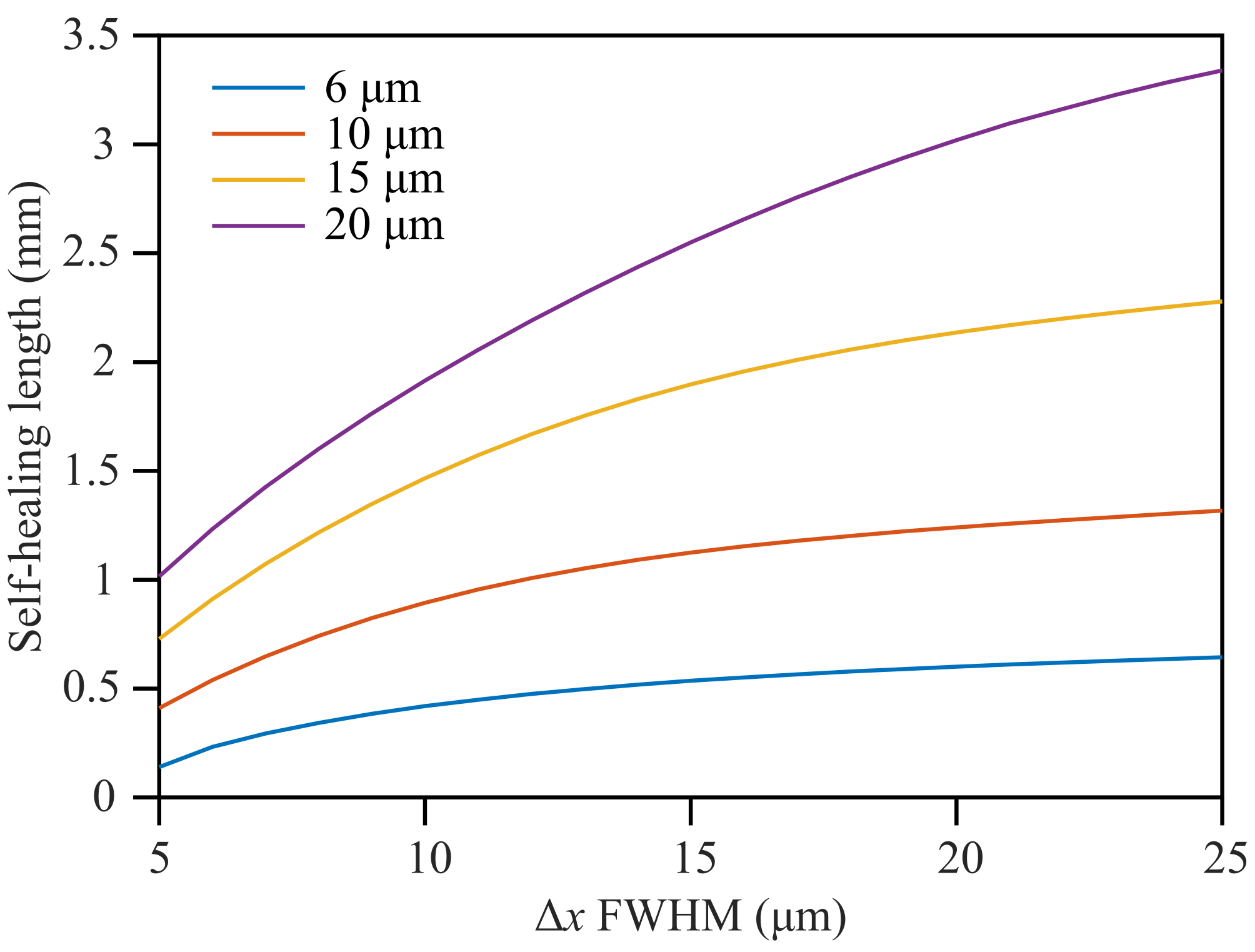}
	\caption{Simulations of the self-healing length for ST light sheets. The four curves correspond to differently sized obstacles placed in the path of the ST light sheet. The self-healing length is plotted against the FWHM with of the light sheet $\Delta x$. See the text for the definition of the self-healing length.} 
	\label{Fig:Theory}
\end{figure}

\begin{figure*}[t!]
	\centering
	\includegraphics[width=17.0cm]{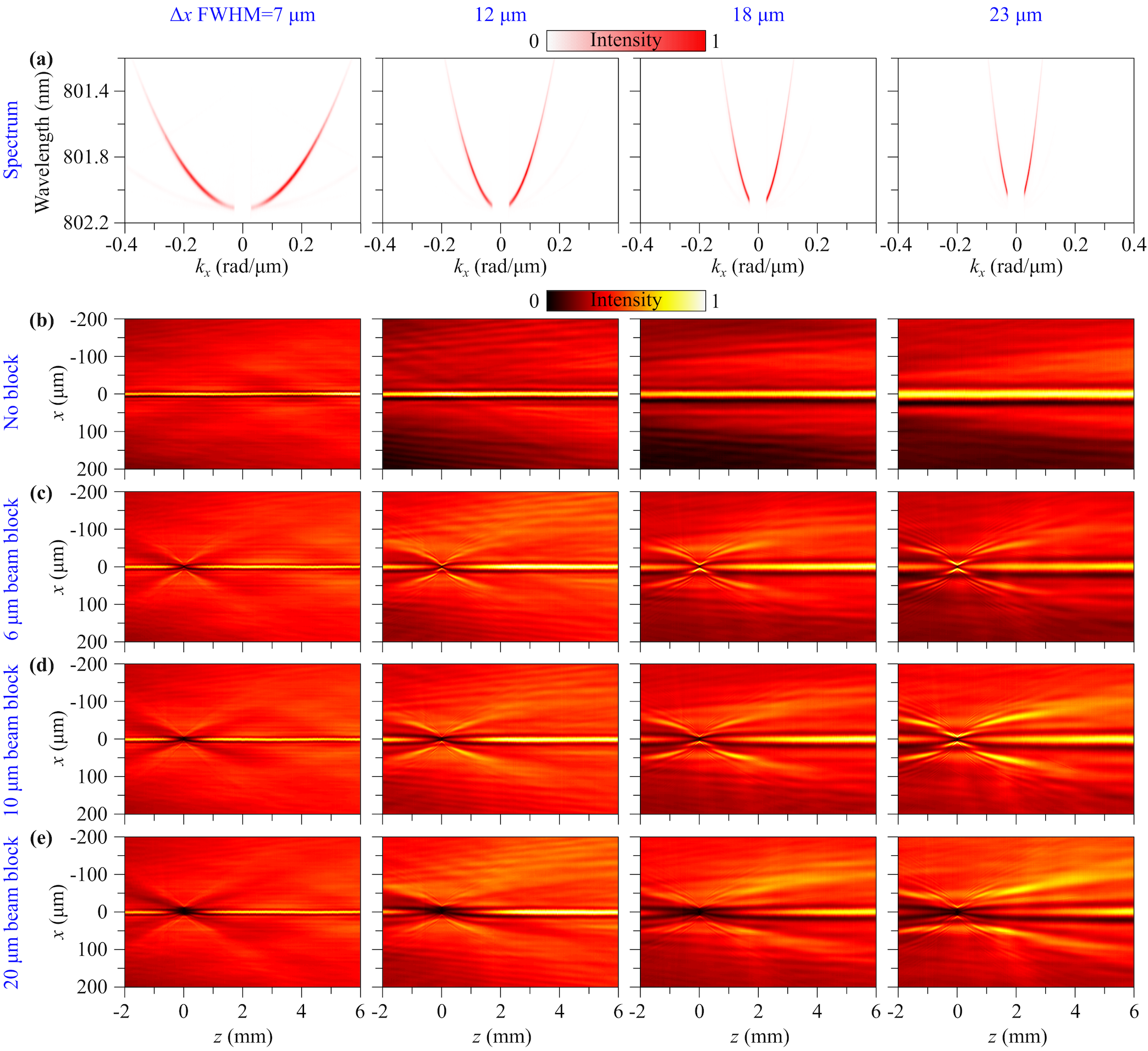}
	\caption{Experimental results confirming the self-healing of ST wave packets after traversing an obstruction. Each column corresponds to a different ST wave packet, all having the same temporal bandwidth $\Delta\lambda\approx0.8$~nm but having different spatial bandwidths $\Delta k_{x}$ and thus different beam FWHM widths $\Delta x=7$, $12$, $18$, and $23$~$\mu$m, respectively. These wave packets correspond to different tilt angles $\theta$ of the spectral hyperplane with respect to the light cone [Fig.~\ref{Fig:Concept}(a)]. (a) Measured spatio-temporal spectral intensity $|\tilde{E}(k_{x},\lambda)|^{2}$, corresponding to $\theta=90^{\circ}$, $53.1^{\circ}$, $48.4^{\circ}$, and $46.8^{\circ}$. (b-e) Measured time-averaged transverse intensity distribution (along $x$; the distribution along $y$ is not shown) at axial positions $-2<z<6$~mm ($z=0$ is the location of the obstruction). Measurements are repeated with obstructions of different widths $d$: (b) no obstacle; (c) $d=6$~$\mu$m; (d) $10$~$\mu$m; and (e) $20$~$\mu$m.}
	\label{Fig:Results} \vspace{-3mm}
\end{figure*}

To identify a self-healing distance, we first define a normalized fidelity of the wave packet after an obstruction with respect to the original ST wave packet. Denoting the time-averaged diffraction-free intensity profile in \textit{absence} of the obstruction $I(x,z)$ and in its \textit{presence} $I_{\mathrm{obs}}(x,z)$ (based on Eq.~\ref{Eq:MainEquation}), we define the fidelity $F(z)$ along the axis as
\begin{equation}
F(z)=\frac{\int dxI(x,z)I_{\mathrm{obs}}(x,z)}{\sqrt{\int dxI^{2}(x,z)}\sqrt{\int dxI_{\mathrm{obs}}^{2}(x,z)}}.
\end{equation}
Of course $F(z)=1$ only when $I_{\mathrm{obs}}(x,z)=\alpha I(x,z)$, i.e., when the field has completely self-healed; $\alpha$ is a real constant less than unity due to finite energy loss at the obstruction. The self-healing distance is defined as the axial distance after which $F(z)>0.9$. We plot in Fig.~\ref{Fig:Theory} the calculated self-healing distance for an ST wave packet using obstacle width $d$ in the range 6 to 20~$\mu$m. We vary the spatial bandwidth of $\Delta k_{x}$ of the ST wave packet to produce transverse widths $\Delta x$ in the range 5 to 25~$\mu$m at a fixed tilt angle $\theta=90^{\circ}$ of the spectral hyperplane plane $\mathcal{P}$ [Fig.~\ref{Fig:Concept}(a)] and assume a spectral uncertainty of $\delta\lambda=25$~pm throughout. We note in all cases that only a few millimeters of axial propagation are required for self-healing; that larger obstacles result in a larger self-healing distance for fixed beam width $\Delta x$; and that narrower beams heal faster than wider ones for the same obstacle width $d$.

We examine in our experiments 4 ST wave packets that all share the same bandwidth $\Delta\lambda\sim1$~nm, but differ in their spatial bandwidth $\Delta k_{x}$ and thus also in their transverse spatial width $\Delta x$: 7, 12, 18, and 23~$\mu$m [Fig.~\ref{Fig:Results}(a)]. Because the spatial and temporal frequencies are correlated, changing $\Delta k_{x}$ while holding $\Delta\lambda$ fixed requires tilting the spectral hyperplane with respect to the light-cone; i.e., changing the angle $\theta$ in Fig.~\ref{Fig:Results}(a). The measured spectral uncertainty is $\delta\lambda\approx25$~pm, which is limited by the resolution of the spectrometer. The calculated value is $\delta\lambda\approx14$~pm based on the illuminated area of the grating. We first examine the spatio-temporal spectra of the ST wave packets by subjecting the field reflecting back from the SLM to a Fourier transform via an appropriate lens arrangement \cite{Kondakci17NP} and record $|\tilde{E}(k_{x},\lambda)|^{2}$ with a CCD camera. The measured spectral intensities are plotted in Fig.~\ref{Fig:Results}(a). The data confirms that by changing $\theta$ from $46.8^{\circ}$ to $90^{\circ}$, the spectral hyperplane intersects with the light-cone in hyperbolas of increasing curvature. Consequently, a smaller $\Delta k_{x}$ (larger $\Delta x$) can be associated with the same $\Delta\lambda$ (Fig.~\ref{Fig:Results}(a) from left to right). We remove the zero-order diffracted component from the SLM using a beam-block in the Fourier domain, which results in the spectral intensity vanishing in the vicinity of $k_{x}$.  

The ST wave packet is observed in physical space after reconstituting the wave packet at the grating by scanning a CCD camera along the $z$-axis on a long translation stage in absence of any obstacles. The evolution of the time-averaged transverse intensity distribution $I(x,z)=\int dt|E(x,z,t)|^{2}$ is thus recorded along $z$, as shown in Fig.~\ref{Fig:Results}(b) for the 4 ST wave packets.

Next we introduce obstructions of width $d=6$, 10, and 20~$\mu$m in the path of \textit{each} ST wave packet using the configuration shown in Fig.~\ref{Fig:Concept}(d). The measurements of these $3\times4=12$ experiments are plotted in Fig.~\ref{Fig:Results}(c)--\ref{Fig:Results}(e). First, in all these experiments, the ST wave packet self-heals after a few-mm of axial propagation, as expected from Fig.~\ref{Fig:Theory}. Secondly, self-healing occurs for \textit{fixed} obstruction width $d$ after a significantly \textit{shorter} distance for ST wave packets with \textit{smaller} transverse size $\Delta x$ (compare the first column to the last in Fig.~\ref{Fig:Results}). This can be explained by noting that the larger $k_x$ components (corresponding to larger propagation angles with respect to the optical axis) involved in the synthesis of beams with smaller $\Delta x$ help more rapidly wrapping the field around the obstacles. Finally, for ST wave packets of fixed $\Delta x$, the self-healing distance increases with $d$ as expected from Fig.~\ref{Fig:Theory}.

In conclusion, we have demonstrated that the recently synthesized class of propagation-invariant ST wave packets or \textit{pulsed} beams exhibit the self-healing characteristics associated with the more familiar class of \textit{monochromatic} diffraction-free beams, such as Bessel or Airy beams. After traversing an obstruction, the scattered component diffracts away in the usual manner whereas the unscattered component that retains the spatio-temporal correlations maintains the propagation-invariant behavior. As such, these ST wave packets maybe useful for time-gated imaging in turbid media or biological samples. Finally, we draw the reader's attention to the fact that the spatio-temporal correlations introduced into the spectrum is the continuous counterpart of the correlations realized between discretized degrees of freedom, such as polarization and spatial modes in \cite{Kagalwala13NP,Okoro17Optica}. In other words, ST wave packets represent optical fields into which continuous `classical entanglement' has been introduced.

\noindent
\textbf{Funding}. U.S. Office of Naval Research (ONR) under contract N00014-17-1-2458.

\noindent
\textbf{Acknowledgments}. We thank M. B. Nasr for preparing the obstructions, and D. Mardani and G. K. Atia for useful discussions.

\bibliography{diffraction}

\end{document}